
\baselineskip=18pt plus 2pt minus 1pt
\magnification=1200
\hsize=5.7truein
\vsize=8.4truein
\voffset=24pt
\hoffset=.1in
\centerline{\bf Is there spin-charge separation}
\centerline{\bf in the 2D Hubbard and ${\rm t-J}$ Models}
\centerline{\bf at low electronic density?}
\vskip .3in
\centerline{Y. C. Chen$^1$, A. Moreo$^2$, F. Ortolani$^3$, E. Dagotto$^2$, and
T. K. Lee$^4$ }
\vskip .15in
{\it \centerline{$^1$Dept. of Physics, National Tsing Hua Univ., Hsinchu,
Taiwan}
\vskip .15in
\centerline{$^2$Dept. of Physics and National High Magnetic Field Lab,}
\centerline{Florida State University, Tallahassee, FL 32306, USA}
\vskip .15in
\centerline{$^3$Departimento di Fisica, Instituto Nazionale di Fisica
Nucleare,}
\centerline{Universita di Bologna, I-40126 Bologna, Italy}
\vskip .15in
\centerline{$^4$Center for Stochastic Processes in Science and Engineering}
\centerline{and Department of Physics,}
\centerline{Virginia Polytechnic Institute and State University}
\centerline{Blacksburg, VA 24061, USA} }
\vskip .3in
The spin and density correlation functions of the
two-dimensional Hubbard model at low electronic density $\langle n \rangle$
are calculated in the ground state by using
the power method, and at finite temperatures by using the quantum Monte Carlo
technique. Both approaches produce similar results, which are in close
agreement with numerical and high temperature expansion results for the
two-dimensional ${\rm t-J}$ model.
Using perturbative approximations, we show that the examination of the
density correlation function alone is not enough to
 support recent claims in the literature
that suggested spin and charge separation in the low electronic density
 regime of the ${\rm t-J}$ model.


\vfil\eject

The normal state of the high temperature
superconductors does not behave as an ordinary Fermi liquid (FL).$^1$
For the last several years, Anderson$^2$ has strongly supported the idea  that
instead it may be described as
a Tomonaga-Luttinger liquid (TLL).$^3$
The TLL state has been shown by exact theories$^4$ and numerical
studies$^{5,6}$
to be the ground state of the Hubbard model, and also the ${\rm t-J}$ model,
in one dimension. A special property
of the TLL state is the separation of spin and charge degrees of freedom.

For these ideas to be applicable to the cuprates, the crucial question
is whether phenomenologically realistic two dimensional models of
correlated electrons present features in the ground state similar to
those of their one dimensional counterparts.
Recently, several numerical studies  have addressed this important
issue, namely the possibility of
spin-charge separation in the two-dimensional ${\rm t-J}$ model. Using
a ground-state projection technique (``power'' method)
to study the low electronic density region,
two of us$^7$ have found indications that spin,
charge and pairing correlations behave in a similar manner as in 1D.
Qualitatively the TLL state
seems to provide a consistent {\it phenomenological} interpretation of
the numerical data. In parallel,
based on the results of high temperature expansions,
Putikka et al.$^8$ have argued that spin-charge separation occurs in the
2D ${\rm t-J}$ model at low and high electronic densities
(particularly at high electronic density).
The gauge theory$^9$ approach also predicts
a non-Fermi liquid (NFL) behavior, although it may not necessarily correspond
to a TLL state.
On the other hand, for the 2D Hubbard model, analytical studies$^{10}$
based upon diagrammatic methods suggest the presence of a Fermi liquid
at low electronic density.

Since it is well known that the ${\rm t-J}$ model is equivalent to the Hubbard
model in
the strong coupling limit, the results described above are apparently
inconsistent. However, the nonperturbative constraint of
no double occupancy in the ${\rm t-J}$ model may produce subtle differences
with the Hubbard model. Precisely, one of the purposes of this paper is
to study numerically the possible variation of physical
properties between these two models as the constraint of no double
occupancy is relaxed. Our analysis shows
that the density and spin correlation functions in the ground state
of the 2D Hubbard and ${\rm t-J}$ models are qualitatively similar at least at
low
electronic density.
To examine the question of spin-charge separation we compare our
results for the density correlations obtained by the power method at zero
temperature on 8$\times$8 and 16$\times$16 clusters,
and by the Quantum
Monte Carlo (QMC) method$^{11}$ at finite temperature,
with that of the high temperature expansions.$^8$ Excellent agreement between
the results of these three methods is obtained.
However, from our analysis
we cannot identify the spinless fermions (SF) Fermi wavevector
2${\bf k}_{F}^{SF}$ as the
characteristic wavevector of the ${\rm t-J}$ and
Hubbard models at low electronic density
as suggested by Putikka et al.$^8$
Both the density and spin correlations can be understood qualitatively
in terms of  perturbative approaches such as
a random-phase approximation (RPA)$^{12}$ with a renormalized
Hubbard coupling ${\rm \bar{U}}$. In addition, the correlations in real space
at low electronic density are shown
to decay so rapidly with distance that the subtle issue of spin-charge
separation in
these models is difficult to address unless these small correlations
are accurately evaluated on a finite cluster, or the functional form of the
correlations at
large distance is obtained with some reliable technique.

The Hamiltonian for the 2D ${\rm t-J}$ model considered here has the form
$$
{\rm H_{tJ}=-t\sum_{<i,j>\sigma} (c^+_{i\sigma} c_{j\sigma}+h.c.)
  +J\sum_{<i,j>}({\bf S}_i\cdot{\bf S}_j-{1\over4} n_in_j),}
\eqno(1)
$$
with the constraint of no double occupancy. The Hamiltonian for the
2D Hubbard model is well-known and will not be reproduced here.
For very large ${\rm U/t}$ the Hubbard model is equivalent to
the ${\rm t-J}$ model with ${\rm J=4t^{2}/U }$, up to two-particle hopping
terms.
The good agreement between results for the ${\rm t-J}$ and Hubbard models
reported below justifies the omission of these three-site terms.$^{13}$
Two numerical methods are used to calculate the equal-time density
and spin correlation functions, ${\rm N({\bf q}) }$ and ${\rm S({\bf q}) }$,
defined by the relations
$$
 {\rm N({\bf q})= \sum_{{\bf r}} e^{i{\bf q}\cdot {\bf r}} }
 <\delta {\rm n_{0}}\delta {\rm n_{\bf r}}>, \eqno(2)
$$
$$
{\rm  S({\bf q})= \sum_{{\bf r}} e^{i{\bf q}\cdot{\bf r}}
 <S_{0}^{z} S_{\bf r}^{z}>,  } \eqno(3)
$$
where ${\rm S_{\bf r}^{z}} = {1\over2}\sum_{\alpha\beta} {\rm c}^+_{{\bf
r}\alpha}
\sigma_{\alpha\beta}^{z} {\rm c}_{{\bf r}\beta}$, and
$\delta n_{\bf r} = \sum_{\sigma}
c^+_{{\bf r}\sigma} c_{{\bf r}\sigma} - \langle n \rangle$. Here
$\langle n \rangle$ is the average
density of electrons.
The brackets in Eqs.(2) and (3) refer to thermal averaging in the grand
canonical ensemble when the QMC method is used.
At zero temperature,
the ground-state wave function  obtained
by the power method in the canonical ensemble
is used to  calculate the average.
We have observed that
the well known fermion determinantal sign problem does not pose
a difficulty in the low-electronic-density region considered in this
paper in any of the techniques. However, this problem becomes more
severe with
increasing density, and thus we restrict our analysis to the
low density region.

The power method has been proven to be very effective
in calculating ground-state correlation functions
of the ${\rm t-J}$ model in 1D$^6$ and 2D.$^7$
The ground state wave function is projected out
by applying a large power of the Hamiltonian, ${\rm (-H)^p}$, to a trial
wave function.
The power p required to reach convergence depends on the choice of the
trial functions.
For the case of the Hubbard model we use the well-known
Gutzwiller wave function,$^{14}$ i.e.
 $\left| GW \right> = g^D \left| FG \right>$, where
$\left| FG \right>$ is the ideal Fermi gas wave function on a lattice
 and D is the total number of doubly occupied sites. Here, $g$ is the
only variational parameter. At $g=0$, the factor $g^D$ becomes
the well known projection operator ${\rm P_d}$ that projects out states
with doubly occupied sites. For the ${\rm t-J}$ model we use the wave function
first proposed by Hellberg and Mele$^{15}$ in 1D and later generalized
by Valenti and Gros$^{16}$ to 2D.
This function, which we shall call HMVG, is basically of the
same form as $\left| GW(g=0) \right>$ i.e. a
Slater determinant for up-spin electrons and
one for down-spin electrons. In addition to these two determinants,
it contains a long range correlation part between all the particles,
$\Pi_{i<j} \left|{\bf r}_i-{\bf r}_j\right|^{\nu}$ (while for nearest-neighbor
particles we chose $\nu=0$). It was shown
in Ref.7 that this wave function is very close to the ground state
for ${\rm J/t\leq2}$ in the low-electronic-density region.

${\rm S({\bf q})}$ and ${\rm N({\bf q}) }$  are plotted in Figs.1a and 1b,
respectively,
against momenta along the $\Gamma$-X-M-$\Gamma$
directions for an 8$\times$8 cluster.
The open circles represent the QMC results obtained at temperature
${\rm T=t/10}$, ${\rm U=4t}$
and ${\rm \langle n \rangle =0.159}$. The open squares are for ${\rm U=8t}$ and
${\rm \langle n \rangle = 0.155}$,
at ${\rm T=t/8}$. The ${\rm U=8t}$ results deviate further from the ideal
Fermi gas shown by the dashed line than the ${\rm U=4t}$ data.
The open triangles represent
 ground-state results for ${\rm U=8t}$
obtained for the same lattice with 10 particles ($\langle n \rangle =
10/64 \approx 0.156$),
 by applying ${\rm p= 12}$ powers of the Hubbard Hamiltonian to the
$\left| GW\right>$ wave
function with $g=0.5$. In most regions of ${\bf q}$-space the open triangles
and squares take the same values. It is gratifying to find out this excellent
agreement between two very different numerical techniques,
namely the power method and QMC.
To gauge the effect of the constraint of no double occupancy, in the
same figures we also
present results for the ${\rm t-J}$ model$^7$ at ${\rm J=0.1t}$.
These results, which are represented
by the solid triangles in Fig.1a-b, are
obtained from the trial wave function HMVG-$\nu=0.1$ with power ${\rm p=16}$.
 The ${\rm t-J}$ model results agree very well
with the open triangles which corresponds to the Hubbard model with
${\rm U=8t}$ (${\rm J=t/2}$ in the ${\rm t-J}$ model language). Hence, there is
little difference in the correlation functions between strong
and intermediate couplings.

Let us now analyze the implications of Fig.1a-b.
All the curves in Fig.1a have peaks at the 2${\bf k}_F$ wavevectors, except
the ideal Fermi gas  which only presents a discontinuous derivative.
The peak size increases with the value of
${\rm U/t}$. The presence of these peaks$^7$ implies a stronger
spin-density-wave correlation at finite coupling than the ideal gas.
It is interesting to notice that a
similar peak is observed$^6$ in 1D. Though the magnitude
of the peak in 1D is much greater than in 2D,
this large difference may be partly due to the
dimensionality effect. Unlike 1D in which there is only one 2${\bf k}_F$
wavevector, in 2D there is a characteristic vector in each direction in
the two dimensional momentum space, each one carrying a peak in ${\rm
S({\bf q})}$.
In contrast to the spin correlations,
${\rm N({\bf q}) }$ shown in Fig.1b is reduced at ${\bf q}=2{\bf k}_F$ as
compared to the values of the ideal Fermi gas shown by the dashed line.
The reduction is larger as the coupling ${\rm U/t}$ is increased. The plateau
observed for $k>2k_F$ for the ideal gas seems to have shifted to
a larger value of $k$. Putikka et al.$^8$ have argued this new
wavevector to be 2${\bf k}_{F}^{SF}$. The presence of the
Fermi wavevector ${\bf k}_{F}^{SF}$ for a spinless fermionic system
would imply the separation of charge and spin degrees of freedom.

To examine the important issue of whether
 2${\bf k}_{F}^{SF}$  appears in the numerical results, and in order
to reduce possible finite size effects
we have calculated ${\rm N({\bf q})}$ on a 16$\times$16 lattice using
both the QMC and power-method techniques.
In Fig.2, ${\rm N({\bf q})}$  is plotted as a function of
${\bf q}$ along the diagonal direction ${\rm q_x=q_y}$.
The open squares represent the QMC results obtained at temperature
${\rm T=t/4}$, ${\rm U=8t}$,
and ${\rm \langle n \rangle = 0.2}$.
The open triangles represent
 ground-state results obtained for the same cluster with 50 particles
(i.e. ${\rm \langle n \rangle \approx 0.195}$)
 by applying twelve powers of the ${\rm t-J}$ Hamiltonian to the HMVG
variational wave
function with $\nu=0.04$. The results of the high temperature expansions$^8$
corresponding to  ${\rm J=t/2}$ and temperature ${\rm T= J/2}$ are indicated
by the solid line. There is very little difference between the results of
these three numerical methods. This implies that the numerical
accuracy of the data is not questionable, but only the interpretation needs to
be
analyzed carefully.
For comparison, the result of the ideal Fermi
gas is plotted as the thin continuous line in Fig.2. It is obvious that the
results of the interactive system deviates appreciably from the ideal
gas results. However,
these results do not unambiguously support
the identification of a
singularity at
2${\bf k}_{F}^{SF}$. A more conservative interpretation is that they
indicate a broad maximum
of ${\rm N({\bf q})}$ at ${\bf q}=(\pi,\pi)$ which may just reflect
the short-range effective repulsion between particles.

To explore further this assumption
we compare ${\rm N({\bf q})}$ against the ${\rm T=0}$
RPA results (dashed line in Fig.2).
The best fit is obtained
by choosing the renormalized
interaction ${\rm {\bar U}}$ to be ${\rm 12t}$. Thus, the apparent shift
of the characteristic wavevector can be mimic very well by a simple
perturbative (Fermi liquid based) approach.
We have also calculated ${\rm S({\bf q})}$
using RPA$^{12}$.
The peaks at 2${\bf k}_{F}$ are also reproduced this time using
a smaller effective coupling ${\rm {\bar U=3t}}$.
Such a
qualitative description of spin and density correlation functions
in terms of a simple RPA tends to support the point of view that
the ground state of the Hubbard model is just a strongly correlated
Fermi liquid. But this may also be misleading. It is also possible to
reproduce the correlation functions of a $one$ dimensional
Hubbard model by using the RPA approximation.
Just like in two dimensions,
a small effective interaction ${\rm {\bar U}}$ is enough to
produce a large peak in ${\rm S({\bf q})}$ at the proper wavevector
2${\bf k}_{F}$. To fit the
density correlation function, a larger ${\rm {\bar U}}$ is needed i.e.
the systematic behavior is very similar in 1D and 2D (as emphasized in
Ref.7).

The excellent agreement between techniques that work at
zero and finite temperature shown in Fig.2 suggests that the shift
in ${\rm N({\bf q}) }$ cannot be due to subtle long distance correlation
functions but to short distance effects. To study this hypothesis we
analyzed in real space the density-density correlation, ${\rm C({\bf r}) }$,
for the case of the one band Hubbard model. Fig.3a shows that this
correlation decays rapidly with distance and it becomes negligible
at four lattice spacings away from the origin (numerically the signal at
this distance is approximately ${\rm 5 \times 10^{-4} C({\bf r}=0)}$).
These 2D correlations are considerably smaller than those obtained
in the case of the one dimensional Hubbard model, which we know shows
spin-charge separation.
This analysis shows that it would be difficult
to obtain reliable numerical information about the behavior of the
correlation functions at distances larger than a few lattice spacings.
Thus, a
proper study  of spin-charge separation seems beyond present
day accuracy of computational and series expansion
analysis at low electronic density.

Can the results of our analysis be extended to higher densities? In
Fig.3b, ${\rm N({\bf q}) }$ is shown at quarter-filling using the
Hubbard model with ${\rm U/t=8}$
and the QMC technique. The results deviate considerably
from the non-interacting Fermi gas, but they can be accurately reproduced
by a simple perturbative calculation (first order) with an effective
coupling ${\rm {\bar U} = 4t}$ (see also Ref.17). Then, we believe that
our conclusions for the Hubbard model can
be extended to the domain $0.0 \leq \langle n \rangle \leq 0.5$.
At higher densities the perturbative approach
breaks down at intermediate and large couplings, due to antiferromagnetism.
On the other hand, the results for the ${\rm t-J}$ model at this density
are very similar to those of non-interacting spinless fermions, as
remarked in Ref.8 using the high temperature expansions. The possible
origin of this discrepancy is currently under study.

In summary, we have presented spin and density correlation functions
for the one band Hubbard model at low electronic
density. The ground-state results obtained by
the power method agree well with the finite-temperature results obtained
by QMC. Measurements at intermediate ${\rm U/t}$ couplings for the
Hubbard model are consistent with
strong coupling data for the ${\rm t-J}$ model. Compared against the
ideal gas results, we confirm that
${\rm N({\bf q})}$ is appreciably reduced at ${\bf q}=2{\bf k}_F$ as claimed by
Putikka et al.$^8$ This difference increases with the
strength of the Hubbard interaction. The enhancement of spin-density-wave
correlation as shown by the appearance
of peaks at the $2{\bf k}_F$ wavevectors
for ${\rm S({\bf q})}$ also increases with ${\rm U/t}$. This result,
first observed in Ref.7, is  confirmed by the present study on larger
clusters and thus finite size effects seem small. On the other hand,
the RPA  approximation can provide a rough qualitative understanding of
all these results. In addition,
examining ${\rm N({\bf q})}$ on a 16$\times$16 lattice we
did not find evidence for the presence of the characteristic wavevector of a
spinless Fermion model. Actually, the density correlations in real space
decay so rapidly that making any statement about their asymptotic
behavior based on numerical techniques at finite temperature is risky.
Thus, based on the current available information it
is $not$ possible to conclude that spin-charge separation takes
place in the low electronic density of the 2D Hubbard and ${\rm t-J}$ models.
However, we cannot rule out this possibility either.
The complete separation of
spin and charge as in the infinite ${\rm U}$ limit of the 1D Hubbard model may
not be
a proper guidance for 2D studies. A possible scenario is that although
charge and spin are separated,
they interact strongly as in the finite ${\rm U/t}$ Hubbard model in 1D. This
question is
currently  being studied.

\medskip
We thank W. O. Putikka for useful conversations and suggestions.
TKL would like to thank the Materials Science Center and Department of
Physics of National Tsing Hua University for their hospitality
during his visit where part of this work is carried out.
This work was partially supported  by
the National Science Council of Republic of China, Grant Nos.
NSC82-0511-M007-140.
A. M. and E. D. are supported by the Office of Naval Research under
grant ONR N00014-93-0495. The numerical QMC calculations were carried out on
the Cray YMP at Florida State University. We thank the Supercomputer
Computational Research Institute for its support.
\medskip

\vfil\eject

\centerline{REFERENCES}
\medskip

\item{1}{\it The Los Alamos Symposium-1989; High Temperature
Superconductivity}, edited by K.S. Bedell, D. Coffey, D.E. Meltzer,
D. Pines, and J.R. Schrieffer (Addison-Wesley, Redwood City, CA,1990).

\item{2}P.W. Anderson, Phys. Rev. Lett. \underbar{64}, 1839 (1990);
{\it ibid}.\underbar{65}, 2306 (1990).
M. Ogata and P.W. Anderson, Phys. Rev. Lett. \underbar{70}, 3087 (1993),

\item{3}S. Tomonaga, Prog. Theor. Phys., \underbar{5}, 544 (1950);
J.M. Luttinger, J. Math. Phys., \underbar{4}, 1154, (1963).

\item{4}F.D.M. Haldane, Phys. Rev. Lett. \underbar{45}, 1358 (1981);
N. Kawakami and S.K. Yang, Phys. Rev. Lett. \underbar{65}, 2309
(1990).

\item{5} M. Ogata et al., Phys. Rev. Lett.
\underbar{66}, 2388 (1991);
H. Shiba and M. Ogata, Prog.
Theor. Phys. Supple.\underbar{108}, 265 (1992).

\item{6}Y.C. Chen and T.K. Lee, Phys. Rev. B\underbar{47}, 11548
(1993);{\it Proceedings of the Beijing International
Conference on High-Temperature Superconductors}, edited by Z. Z. Gan,
S.S. Xie, and Z.X. Zhao, 829 (1993), World Scientific, Singapore.

\item{7}Y.C. Chen and T.K. Lee, to appear in Z. Phys. B.

\item{8}W.O. Putikka, R.L. Glenister, R.R.P. Singh and H. Tsunetsugu,
unpublished.

\item{9}P.A. Lee and N. Nagaosa, Phys. Rev. B\underbar{46}, 5621 (1992).

\item{10}J.R. Engelbrecht and M. Randeria, Phys. Rev. B\underbar{45}, 12419
(1992).

\item{11} For references see E. Dagotto, preprint (to appear in
Rev. Mod. Phys.).

\item{12}N. Bulut, D.J. Scalapino adn S.R. White, Phys. Rev. B\underbar{47},
2742 (1993).

\item{13} Putikka et al$^8$ have
neglected the $-{{1}\over{4}} {\rm n_i n_j}$ term Eq.(1).

\item{14}M. C. Gutzwiller, Phys. Rev. Lett. \underbar{10}, 159 (1963).

\item{15}C. Stephen Hellberg and E.J. Mele, Phys. Rev. Lett. \underbar{67},
2080 (1991).

\item{16}R. Valenti and C. Gros, Phys. Rev. Lett. \underbar{68}, 2402
(1992).

\item{17} A.-M. Dar\'e, L. Chen, and A.-M. S. Tremblay, Phys. Rev.
B\underbar{49}, 4106 (1994).

\vfil\eject

\centerline{Figure Captions:}
\medskip
\item{Fig. 1} (a) Spin correlation function ${\rm S({\bf q})}$ and (b)
density correlation function ${\rm N({\bf q})}$
in momentum space along the $\Gamma$-X-M-$\Gamma$ directions on a
$8\times8$ square lattice. Open circles represent QMC results at ${\rm U=4t}$,
${\rm \langle n \rangle= 0.159}$ and ${\rm T=t/10}$.
The open squares correspond to ${\rm U=8t}$,
${\rm \langle n \rangle = 0.155}$ and ${\rm T=t/8}$.
The open triangles represent ground state results
for the Hubbard model at ${\rm U=8t}$ with 10 electrons. The solid triangles
denote
ground-state results for the ${\rm t-J}$ model at ${\rm J=0.1t}$. The
dashed line is the result for an ideal Fermi gas.

\item{Fig. 2} The density correlation ${\rm N({\bf q})}$ along
the diagonal direction ${\rm q_x=q_y}$ on
a $16\times16$ cluster.  The open squares represent the QMC results
obtained at ${\rm T=t/4}$, ${\rm U=8t}$ and ${\rm \langle n \rangle
= 0.2}$ (results on $12\times12$ lattices are also shown). The open
triangles are power method
ground-state results for the ${\rm t-J}$ model at ${\rm J=t/2}$. The solid line
is the result obtained by the high temperature expansion of Ref.8.
The dashed line denotes the RPA prediction.
The result for an ideal Fermi gas is plotted as a thin continuous line.

\item{Fig. 3} (a) Density-density correlation ${\rm C({\bf r}) = \langle
n_{\bf 0}
n_{\bf r} \rangle - \langle n \rangle^2 }$ as a function of the distance
${\bf r}$.
The solid line is the power method result obtained at ${\rm J/t=0.5}$
(i.e. ${\rm U/t=8}$ in the Hubbard model) on a 16$\times$16 cluster
and density $\langle n \rangle \approx. 0.20$. The dashed line denotes the
result for a tight-binding non-interacting system ${\rm U/t=0}$ of the
same size. The
correlations are considered along the diagonal of the lattice (the lattice
spacing is equal to one);
(b) The density correlation ${\rm N({\bf q})}$ along the direction
$\Gamma$-X-M-$\Gamma$ using the QMC technique (full squares) on a $16 \times
16$
cluster, at ${\rm T=t/4}$, ${\rm U/t=8}$, and density $\langle n \rangle
= 0.5$. The dashed line indicates the result for an ideal Fermi gas,
while the dotted line corresponds to a perturbative calculation using
an effective coupling ${\rm {\bar U}=4t}$. The continuous line
corresponds to non-interacting spinless fermions at the same density and
temperature.

\end